\documentclass[pre,twocolumn,superscriptaddress]{revtex4-1}
\usepackage[pdftex]{graphicx}
\usepackage{amsmath}
\usepackage{color}
\newcommand{\beq}{\begin{equation}}
\newcommand{\eeq}{\end{equation}}
\newcommand{\bea}{\begin{eqnarray}}
\newcommand{\eea}{\end{eqnarray}}

\newcommand{\avg}[1]{\langle{#1}\rangle}
\newcommand{\Avg}[1]{\left\langle{#1}\right\rangle}

\newcommand{\gin}[1]{}
\newcommand{\luca}[1]{}

\begin{document}
\title{Condensation and topological phase transitions in a dynamical network model  \\with rewiring of the links}

\author{Luca Ferretti} 
\thanks{These authors contributed equally.}
\affiliation{Syst\'ematique, Adaptation et Evolution (UMR 7138), UPMC Univ Paris 06, CNRS, MNHN, IRD, Paris, France}
\affiliation{CIRB, Coll\`ege de France, Paris, France}

\email[Email:]{luca.ferretti@gmail.com}
\author{Marcello Mamino}
\thanks{These authors contributed equally.}
\affiliation{CMAF, Universidade de Lisboa, 1649-003 Lisboa, Portugal}
\author{Ginestra Bianconi}
\affiliation{School of Mathematical Sciences, Queen Mary University of London, London E1 4NS, United Kingdom}

\begin{abstract}
Growing network models with both heterogeneity of the nodes and topological constraints can give rise to a rich phase structure. 
We present a simple model based on preferential attachment with rewiring of the links. Rewiring probabilities are modulated by the negative fitness of the nodes and by the constraint for the network to be a simple graph. At low temperatures and high rewiring rates, this constraint induces a Bose-Einstein condensation of paths of length 2, i.e. a new phase transition with an extended condensate of links. The phase space of the model includes further transitions in the 
size of the connected component and the degeneracy of the network.    
\end{abstract}
\maketitle

\section{Introduction}
 
Over the last ten years, critical phenomena  \cite{Doro_crit,Vespignani_book}  in complex networks  \cite{Latora,Doro_book} have attracted large attention. Among the different classes of critical phenomena, condensation transitions play a special role. In fact in this case it is the structure of the network itself that changes topology across the phase transition.
Condensation phenomena  occur in a network when a finite fraction of a special type of subgraph (link, triangle ect.) is found in a relatively small region of the network.  Advances in understanding condensation phase transitions can
help to characterize novel stylized models that present new challenges in the statistical mechanics
of complex systems.
Condensation transitions indeed occur not only in network models but   have been also identified  in other complex systems models such as stylized traffic models, know as the zero-range process \cite{Evans, ball_in_box,Noh1,Noh2, Waclaw}, wealth distribution models \cite{wealth,Mezard}, and models of biological evolution \cite{Kingman, Kadanoff,NonNeutral}.
 
In the context of network theory, condensation phase transition are found in equilibrium network models \cite{Strauss, Burda_strauss1,Burda_strauss2,Doro_cond} as well as in non-equilibrium network models \cite{Bose, Weighted,Redner1, Redner2,fitness,ferretti2012features}.
The first condensation transition in network model  was discovered in 1986 by Strauss \cite{Strauss} in the framework of an equilibrium (Hamiltonian) model of  networks. More recently, a condensation transition that can be mapped to a Bose-Einstein condensation  has been found in the non-equilibrium scale-free growing network model called the  fitness Bianconi-Barab\'asi model \cite{Bose}.
In this model of a growing network, every node $i$ is assigned a fitness $\eta_i$ from a fitness  distribution $\rho(\eta)$ that describes the ability of a node of attracting new links. At each time a new node is added to the network. This node has $m$ links and these  links are attached with higher probability to high fitness-high connectivity nodes following a generalized ``preferential attachment'' rule \cite{BA}.  The resulting network is scale-free \cite{fitness}.
When the fitness  of the fittest node of the network is significantly higher than the mean fitness, 
this node grabs a finite fraction of all the links of the network. This phase transition can be mapped to the Bose-Einstein condensation in a Bose gas.

In this paper we will  characterize the  condensation phase transition in a non-equilibrium growing network model with rewiring of the links. 
Each  node $i$ of the network is  assigned a negative fitness $\xi_i$  from a distribution $\rho(\xi)$ which is a quenched variable. The rewiring process tends to remove links from nodes that have a high negative fitness in this way strongly optimizing the network structure. This model might represent a social network model where agents tend to explore the network and look for good friends (for example, cooperators) with low negative fitness $\xi$ (where $\xi_i$ might indicate the probability of node $i$ to defect). In such network the punishment of high negative fitness nodes (defectors) is dynamically reached in the network through the process of rewiring of the links.
Here, we don't focus on a specific mechanism driving the rewiring process, but we investigate in a simple mathematical framework the effect that the rewiring process can have on the structure and the evolution of growing networks.

As a function of a parameter $T$ that changes the distribution of the negative fitness, and the rewiring rate $r$, we observe a condensation of the links with formation of an ``extended'' condensate, such that for $T<T_c(r)$  or equivalently $r>r_c(T)$ an infinitesimal fraction of the nodes grabs a finite fraction of all the links in the network. 
This appears as ``condensation'' of paths of length $2$ from a single node, i.e. condensation of links on the nearest neighbours of the node. 
This phase transition is a consequence of the fact that we allow only for the generation of simple networks, i.e. we do not allow the presence of  selflinks/tadpoles and multiple links. 
Interestingly, the condensation transition can be mapped  to a Bose-Einstein condensation in a Bose gas. 
The condensation transition impacts also the scaling of the connected component of the network, which becomes sublinear in time in the condensed phase. 
Furthermore, there appears to be another topological transition inside the condensed phase, with the formation of an highly connected core. 

The paper is structured as follows: In section II we describe and solve the model, in section III we show that the model undergoes a condensation transition, in section IV we describe the transitions at $T=0$ where they can be characterized analytically, in section IV we discuss the topological transitions, finally in section V we give the conclusions.


\section{The model: networks with rewiring and negative fitness}
We consider a network growth through preferential attachment and rewiring. 
We start from a finite network of $N_0>mr$ nodes where $m>1$ and $r$ 
are two parameters of the model that we discuss in the following.
New nodes with $m$ links each are added at constant rate $r_{new}=1$ to the system. 
We assume that  the new  links are  attached to an existing  node $i$ with the  ``preferential attachment'' probability \cite{BA}
\beq
\Pi_+(i)=\frac{k_i}{\sum_l k_l}
\eeq
where $k_i$ indicates the degree of node $i$.
Moreover, rewiring of links takes place at rate $mr$. Each node has a ``negative fitness''  $\xi$, randomly extracted from a distribution $\rho(\xi)$. Node that have a high negative fitness $\xi$ are more likely to lose their connections due to the rewiring process. Indeed we assume that each rewiring event corresponds to a link that is removed from node $i$ according to the probability
\beq
\Pi_-(i)=\frac{\xi_ik_i}{\sum_l \xi_lk_l}\label{probm}
\eeq
and is reattached to node $j$ according to $\Pi_+(j)$. 
Note that in this network model we consider only rewiring process that keep the network a simple graph, therefore the reattachments that lead to multiple links and to tadpoles are excluded.
We observe here that given a nonzero minimum $\xi_{min}>0$ of the support of $\rho(\xi)$, it is always possible to rescale all values of $\xi$ such that $\xi_{min}=1$.
We will be mostly interested in models where the rewiring process is more important than the growth process in assigning links, that is, $r>1$. 

As long as the degree of each node is an infinitesimal fraction of all the links, we can write down effective continuum equation which neglect the occurrence of multiple links and tadpoles.
As we will see below in the treatment of the model this assumption is valid asymptotically for large network sizes, as long as the network is not in its condensed phase. 
\subsubsection{Case of homogeneous fitness distribution}
First we solve the degree distribution for the model in which all the negative fitnesses are equal, that is, $\xi_i=1$. 
The continuum equation for the degree of a node is simply ${dk}/{dt}={k}/{2t}$, that is, the same equation as in the Barab\'asi-Albert model \cite{BA}. The birth rate of nodes is also the same, therefore the continuum approach predicts that the degree distribution is the given by the  power law $p(k)\propto k^{-3}$ already found in Barab\'asi-Albert model  with preferential attachment. 
However rewiring enhances the stochasticity of the process, potentially affecting the distribution. As a simple check for the validity of the mean-field approach just outlined, we estimate the variance of the growth of the degree as the sum of the (Poisson) variances $m(1+r)\Pi_+=(1+r)k/2t$ for link attachment and $mr\Pi_-=rk/2t$ for link removal, and promote the above equation to a Langevin equation 
\beq
\frac{dk_i}{dt}=\frac{k_i}{2t}+\sqrt{\frac{(1+2r)k_i}{2t}}\eta_i(t)
\eeq
 with $\eta_i(t)$ a white Gaussian noise with variance 1. 
 Therefore we  obtain   the integrated Fokker-Planck equation 
\beq
\frac{\partial p(k,t)}{\partial t}=-\frac{p(k,t)}{t}-\frac{1}{2t}\frac{\partial (k p(k,t))}{\partial k}+\frac{1+2r}{4t}\frac{\partial^2 (kp(k,t))}{\partial k^2}
\eeq
defined for the  average degree distribution $p(k,t)=\int_1^tdt_iP(k,t|t_i)/t$, where $P(k,t|t_i)$ is the distribution for a node born at time $t_i$. The (normalizable) stationary solution of this equation is 
\beq
p(k)\propto \frac{1}{k^{3}}-3(k+1+2r)
\int_k^\infty dy\frac{ e^{-\frac{2(y-k)}{1+2r}}(2y-1-2r)}{y^4(y+1+2r)^2}
\eeq 
which goes as $p(k)\sim k^{-3}+3(1+2r)k^{-4}+o(k^{-4})$, thereby confirming the validity of the continuum approach for $k\gg r$. The full degree distribution of these models will be presented elsewhere. 


\subsubsection{Case of heterogeneous fitness ditribution}
In the following we will consider the more interesting case in which the  negative fitness of the nodes is not homogeneous over the nodes of the network. 
In order to solve the mean-field continuum equation for the model, we  make the self-consistent assumption that   the denominator of Eq. (\ref{probm}) scales linearly with time, i.e.
\beq
Z=\sum_i\xi_{i}k_i=mCt+o(t),\label{eqcdef}
\eeq
where we call $Z$ the partition function of the model.
The continuum equations, always neglecting occurrences of multiple links and tadpoles, is  asymptotically given by 
\beq
\frac{dk_i}{dt}=\frac{(1+r)k_i}{2t}-\frac{r\xi k_i}{Ct}\quad\Rightarrow\quad k_i(t)=m\left(\frac{t}{t_i}\right)^{(1+r)/2-r\xi/C},\label{eqk}
\eeq
where $t_i$ is the time at which node $i$ is added to the network.
Only  the nodes with increasing average degree 
 $dk/dt>0$ contribute to the tail of the degree distribution.
Then the asymptotic degree distribution for large $k$ is a sum of power laws
\beq
p(k)= 
\int_0^{\frac{(1+r)C}{2r}}  \frac{2C\rho(\xi) \ d\xi}{m((1+r)C-2r\xi)} \left(\frac{k}{m}\right)^{-1-\left[(1+r)/2-r\xi/C\right]^{-1}}\label{eqp}
\eeq
similarly to \cite{fitness,quantum}. Therefore the resulting network is scale-free and develops a power-law tail. The exponent is parametrized by the constant $C$ defined in Eq. $(\ref{eqcdef})$ that we can rewrite as 
\beq
C=\lim_{t\to \infty} \frac{Z}{mt}=\lim_{t\to \infty }\frac{\avg{Z}}{mt}
\eeq
where $\Avg{\ldots}$ indicates the average over the distribution $\prod_{i=1}^t\rho(\xi_i)$.
By inserting the expression $k_i(t)$ given by Eq. $(\ref{eqk})$ in Eq. $(\ref{eqcdef})$ it is easy to show that $C$ is given by 
\beq
r=\int d\xi \rho(\xi)\frac{1}{1-\frac{C(r-1)}{2r\xi}}
\label{self}
\eeq 
The topology of the network will be discussed in section \ref{top}. Here we just note that the network contains essentially a single connected component plus a number of isolated nodes. In fact, a node that loses all links will remain isolated forever, while a connected component separated from the biggest one has a low probability of be generated by rewiring 
and a lifetime of order of the inverse of its fraction of links. 
Moreover, in the normal phase, the network has no clustering in the thermodynamic limit, i.e. its clustering coefficient decreases as an inverse power of time  \cite{ferretti2012features}.

\section{Bose-Einstein phase transition}
Depending on the heterogeneities present in the fitness distribution, the model might undergo a condensation phase transition that can be mapped to a Bose-Einstein condensation in a Bose gas.
In order to show this result it is useful to parametrize the value of the negative fitness by  a ``temperature'' parameter $T$ by defining 
\beq
 \xi_i=e^{\beta\varepsilon_i}
 \label{eps}
 \eeq with  $\varepsilon_i\in[0,+\infty)$ and with   a corresponding distribution $g(\varepsilon)$. The inverse temperature $\beta=1/T$ is a dummy variable that can  change its value from $\beta=0$ to $\beta=\infty$. The high-temperature limit $\beta\rightarrow 0$ corresponds to the neutral model in which all the negative fitness are the same,  discussed in the previous section. 
As mentioned before, this parametrization allows us to find a Bose-Einstein condensation phase transition in this system. Using Eq. $(\ref{eps})$ we can derive an expression for the partition function $Z$ ,i.e. $Z=\sum_i e^{\beta\varepsilon_i}k_i$ with $C=\lim_{t\to \infty }Z/(mt)$. The parameter $C$ is given by Eq.  $(\ref{self})$ which now reads,
\beq
C=\frac{C}{r}\int d\varepsilon \frac{g(\varepsilon)}{1-\frac{C(r-1)}{2r}e^{-\beta\varepsilon}} \label{eqc}
\eeq
In the stationary case Eq. (\ref{eqc}) is exact, even if it is obtained in the continuum approximation, because it depends only on the average degree of the nodes, which is described precisely by the continuum Eq. (\ref{eqk}). For consistency we must impose in Eq. (\ref{eqc}) the constraints $C>2$ and $(r-1)/2r\leq C^{-1}$ to avoid superlinear growth of the degree of nodes with $\varepsilon\rightarrow 0$. 

For low rates of rewiring $r\leq 1$, the integral in the r.h.s of Eq. (\ref{eqc}) is a Fermi-like integral and Eq. (\ref{eqc}) always admits a solution that satisfies this inequality \cite{quantum}. Therefore, the degree distribution of the model follows the expression (\ref{eqp}).

This is also true for the fast rewiring case 
where $r>1$ if the energy distribution $g(\varepsilon)$ satisfy $g(0)>0$, i.e. if in the system there is a finite fraction of nodes with lowest energy $\varepsilon$. 

Instead, in the interesting case $r>1$ and $g(0)=0$, the Eq. (\ref{eqc}) does not always admit solutions. In fact, rearranging Eq. (\ref{eqc}) and defining  the chemical potential $\mu\leq 0$ as $e^{\beta\mu}=(r-1)C/2r$, we obtain the equivalent consistency equation
\beq
r-1=\int d\varepsilon \frac{g(\varepsilon)}{e^{\beta(\varepsilon-\mu)}-1} 
\label{boseint}
\eeq
which  is precisely the usual Bose integral related to the conservation of the number of particles in a Bose gas as long as we map the constant $r-1>0$ to the density of particles in the Bose gas.

Therefore we can predict that a phase transition similar to  Bose-Einstein condensation occurs in our network model with rewiring of the links when the condition
\beq
r-1\leq \int d\varepsilon \frac{g(\varepsilon)}{e^{\beta\varepsilon}-1} 
\label{bect}
\eeq
cannot be satisfied anymore.
As long as $g(\varepsilon) \to 0$ for $\varepsilon \to 0$ we have that the integral on the left hand side of the Eq. $(\ref{bect})$ becomes  smaller while lowering the temperature. Therefore for $\beta$ higher than a critical value $\beta_c$ the Eq. $(\ref{bect})$  cannot be anymore satisfied and we have to assume that some form of condensation of links is taking place.

\begin{figure}
\begin{center}
\includegraphics[width=\columnwidth]{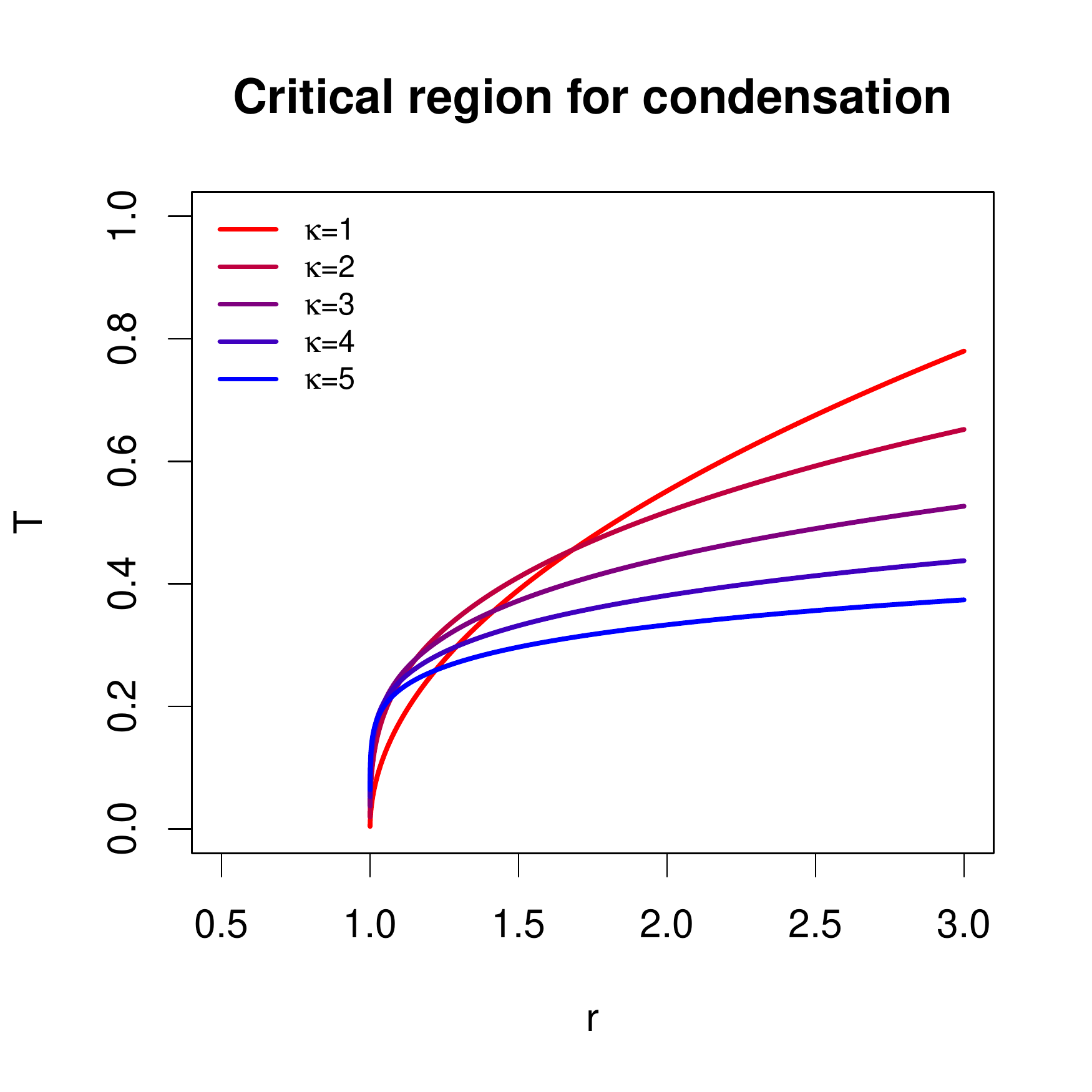}
\end{center} 
\caption{(Color online) Border of the critical region for condensation, according to Eq. (\ref{eqbrt}) for different values of $\kappa$.}\label{fig_crit}
\end{figure}

Note that in this model there are two possible ways to reach the transition: fixing $r>1$ and decreasing $\beta$ beyond $\beta_c(r)$, or fixing $\beta<+\infty$ and increasing the rewiring rate $r$ beyond $r_c(\beta)$
\beq
r_c(\beta)=1+\int \frac{d\varepsilon g(\varepsilon)}{e^{\beta\varepsilon}-1}.
\eeq
For a probability $g(\varepsilon)=(\kappa+1)\varepsilon^{\kappa}$ the border of the critical region, i.e. the critical temperature $T_c=1/\beta_c$ and the critical rate $r_c$ of rewiring are given by 
\beq
\frac{(r_c-1)}{ T_c^{\kappa+1}}=(\kappa+1)\int dx \frac{x^{\kappa}}{e^x-1}\label{eqbrt}
\eeq
The $r-T$ phase plot for the phase transition is shown in Figure \ref{fig_crit}.

\section{Condensation of paths of length 2}

If we would allow for tadpoles and multilinks in the system, links would tend to form a Bose-Einstein condensate on one of the nodes of lower energy, similarly to what happens in the Bianconi-Barabasi fitness model \cite{Bose}. 
However, the constraint that the network is a simple graph changes the nature of the condensate.

We define the quantity $P_{2,i}$ as the number of paths of length 2 starting from a given node $i$.
In terms of the adjacency matrix $a_{ij}$ of the network, the quantity $P_{2,i}$ can be written as
\beq
P_{2,i}=\sum_{j_1=1}^N\sum_{j_2\neq i} a_{i,j_1}a_{j_1,j_2}.
\eeq
Since a path of length 2 starting from node $i$ is given by $a_{i,j_1}a_{j_1,j_2}\leq a_{j_1,j_2}$ with $j_2\neq i$, we have 
\beq
P_{2,i}\leq \sum_{j_1=1,N}\sum_{j_2=1,N}a_{j_1,j_2}=2mt
\eeq
where $mt$ is the number of links in the network.
 The quantity $P_{2,i}$ has another interpretation: is the total number of links connected to the nearest neighbours of node $i$, but not to $i$ itself.
We denote by $P_2=\max_{i}{P_{2,i}}$ and by $p_2$ the following expression
\beq
p_2=\max_{i}\frac{P_{2,i}}{2mt}=\frac{P_{2}}{2mt},
\eeq
which is always smaller or equal to one.

In the normal phase, $p_2$ is infinitesimal for large $t$, because from Eq.(\ref{eqp}) it takes a finite fraction of nodes to have a finite fraction of links, but all nodes have an infinitesimal fraction of nearest neighbours. 


In order to show that there is condensation of $p_2$, we rediscuss the evolution of the node degree as follows. In the step of addition of new nodes, the node gains $m$ links with probability $\frac{k}{2mt}$ each, growing at a rate $m {k}/{2mt}$. In the rewiring step, there are $mr$ rewiring events that could remove links with probability ${\xi k}/{mCt}$, so it loses $mr{\xi k}/{mCt}$ links on average. On the other hand, the rate of links acquired through rewiring is $mr {k}/{2mt}\cdot \left(1-p^{(self)}_i-p^{(mult)}_i\right)$ where $p^{(self)}_i$ and $p^{(mult)}_i$ are the probability of forming selflinks or multilinks attached to node $i$.
Therefore the continuous rate equation for the degree of a generic node $i$ is given by 
\bea
\frac{dk_i}{dt}=\frac{k_i(1-r\xi_i)}{2t}+\frac{rk_i}{2t}\left(1-p^{(self)}_i-p^{(mult)}_i\right)\label{eqdy}
\eea
The probability of selflinks is equal to the probability that the other end of the removed link is actually attached to the node, so it can be obtained by summing the removal probability $\frac{\xi}{mCt}$ over all the $k$ nearest neighbours of the node. This gives 
\bea p^{(self)}_i=\frac{\langle\xi_{NN}\rangle_i k_i}{mCt} \eea
where $\langle\xi_{NN}\rangle_i$ is the average fitness among nearest neighbours of node $i$. Similarly, multiple links are generated by the rewiring of the ends of paths of length 2 from the node, so the probability of multilinks is given by the sum of ${\xi}/{mCt}$ over all the $P_{2,i}$ ends of the paths of length 2 starting from node $i$. This implies
 \bea p^{(mult)}_i=\frac{\langle\xi_{NP_2}\rangle_i k_i}{mCt},\eea
 where $\langle\xi_{NP_2}\rangle_i$ is the average fitness among nodes at the end of paths of length 2 starting from node $i$, weighted by the number of such paths. 
Putting everything together, the final equation for the node $i$ is given by :
\beq
\frac{dk_i}{dt}=\frac{k_i}{2t}+\frac{rk_i}{2t}\left(1-\frac{\langle\xi_{NN}\rangle_i k_i}{mCt}-\frac{\langle\xi_{NP_2}\rangle_i P_{2,i}}{mCt}\right)-\frac{r\xi_i k_i}{Ct}\label{eqkp2}
\eeq
where 
\bea
\langle\xi_{NN}\rangle_i&=&\sum_j a_{ij}\xi_j/k_i\nonumber \\ 
\langle\xi_{NP_2}\rangle_i&=&\sum_{j_1}\sum_{j_2\neq i}a_{i,j_1}a_{j_1,j_2}\xi_{j_2}/P_{2,i}.
\eea
Therefore the continuum equation  for a node $i=i_0$ of minimum energy with $\mu=0$, forbidding selflink and multilinks, is
\beq
\frac{dk_{i_0}}{dt}=\frac{k_{i_0}}{t}-\frac{rk_{i_0}}{2t}\frac{\langle\xi_{NN}\rangle_{i_0} k_{i_0}+\langle\xi_{NP_2}\rangle_{i_0} P_{2,i_0}}{2mrt/(r-1)},\label{keq}
\eeq
where from now on  we will omit the label $i_0$ and we will indicate $P_{2,i_0}$ with $P_2$, $k_{i_0}$ with $k$ and $\Avg{\xi_{NN}}_{i_0}$ and $\Avg{\xi_{NP_2}}_{i_0}$ with $\Avg{\xi_{NN}}$ and $\Avg{\xi_{NP_2}}$ respectively.

If we we assume that $\langle\xi_{NN}\rangle$ and $\langle\xi_{NP_2}\rangle$ are constant in time, it is possible to solve Eq. (\ref{keq}) with $P_2=0$. In this case $k$ would grow almost linearly in time:
\beq
k(t)_{P_2=0}\simeq\frac{k_0t}{t_0+\frac{k_0(r-1)\langle\xi_{NN}\rangle}{2}\log(t/t_0)}\sim \frac{t}{\log(t)} 
\eeq
Both  $\langle\xi_{NN}\rangle$ and $\langle\xi_{NP_2}\rangle$ are  bounded between 1 and $\langle\xi\rangle$, therefore their time dependence does not affect the scaling with time of $k(t)_{P_2=0}$.
It is easy to see that even for $P_2>0$, $k(t)$ would follow the same scaling $k(t)\sim t/\log(t)$ provided that $P_2$ grows not faster than $k$, i.e not faster than $t/\log(t)$. 

\begin{figure}
\begin{center}
\includegraphics[width=\columnwidth]{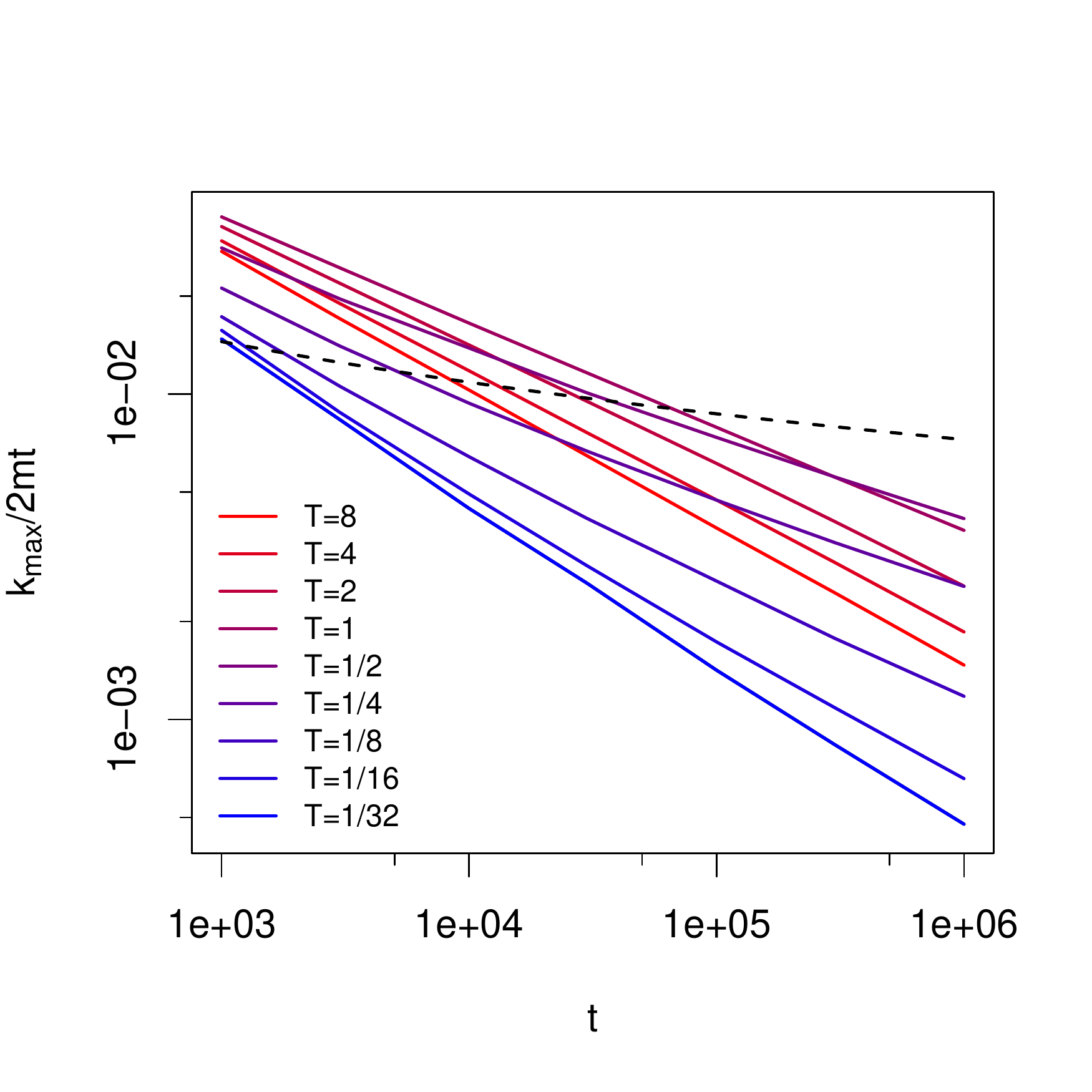}
\end{center} 
\caption{(Color online) Log-log plot of $k_{max}/2mt$ as a function of $t$ in a model with $m=2$, $r=2$, $g(\varepsilon)=3\varepsilon^2$ for different temperatures below and above the critical temperature $T_c\sim 0.5$, averaged over 100 runs. For comparison, the dashed line corresponds to $k_{max}/2mt=0.025/\log(t)$.}\label{kmax}
\end{figure}

\begin{figure}
\begin{center}
\includegraphics[width=\columnwidth]{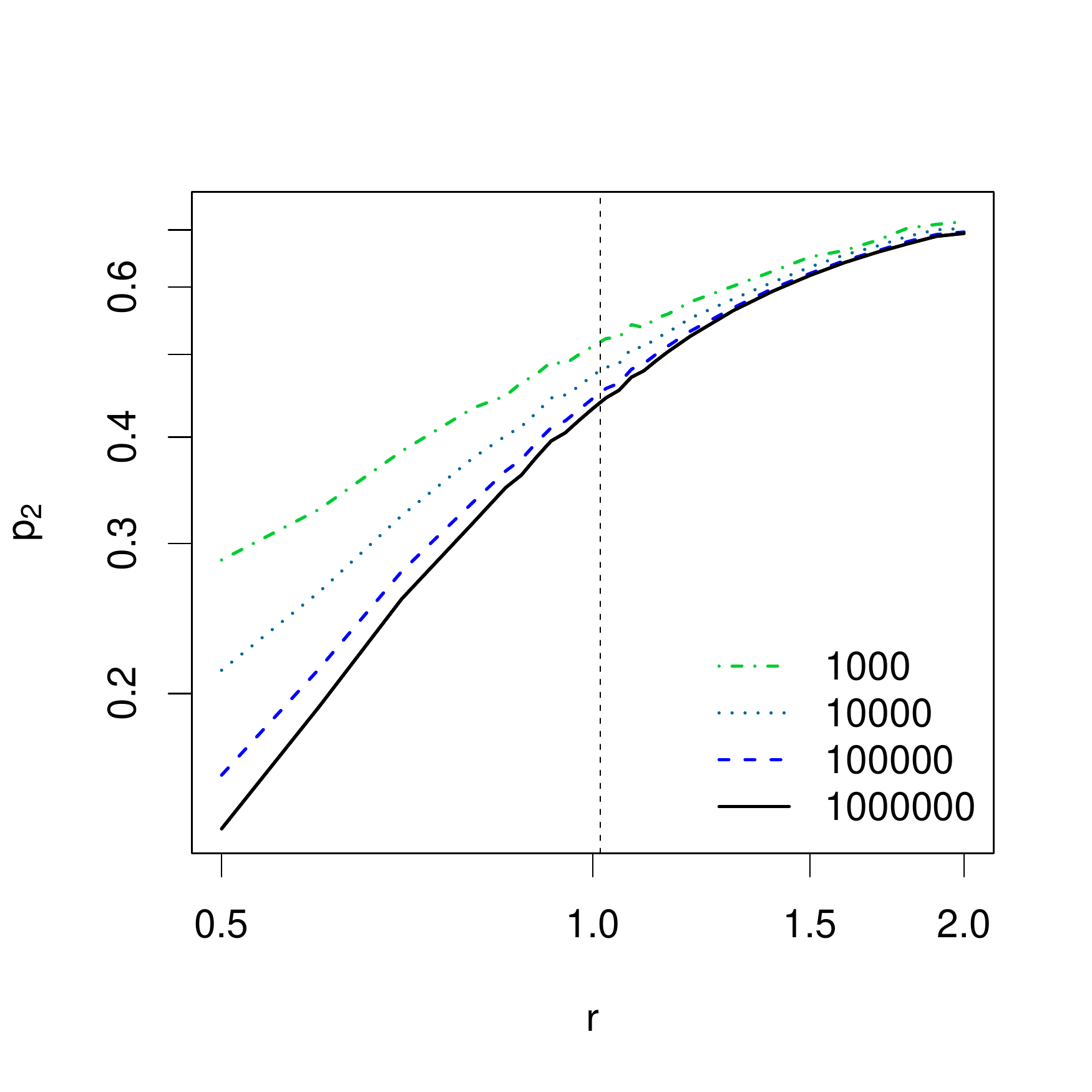}
\end{center} 
\caption{(Color online) Plot of $p_2$  as a function of $r$ for different times from $10^3$ to $10^6$ in a model with $m=2$, $T=0.125$, $g(\varepsilon)=3\varepsilon^2$, averaged over 100 runs. 
}\label{p2_max}
\end{figure}

\begin{figure}
\begin{center}
\includegraphics[width=\columnwidth]{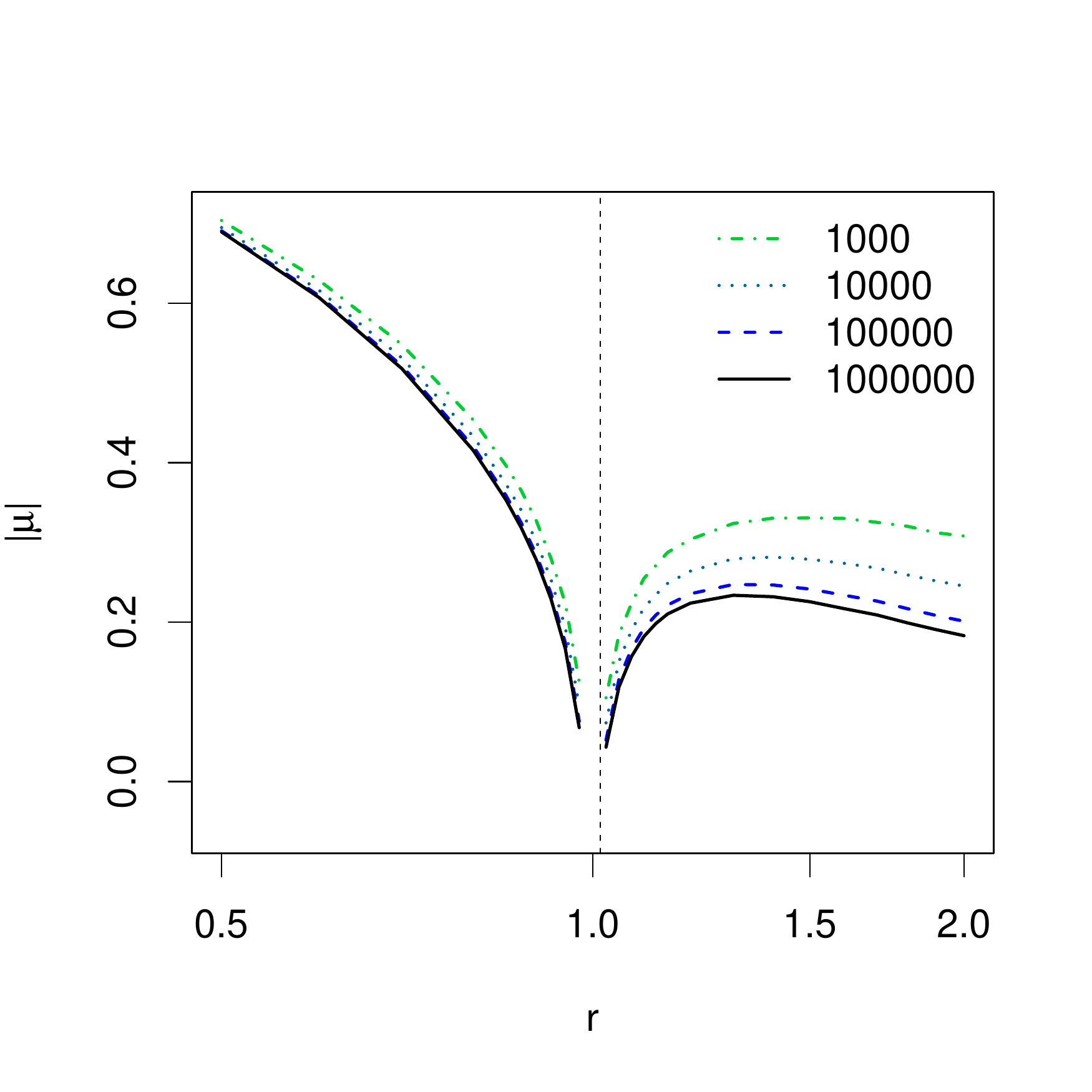}
\end{center} 
\caption{(Color online) Plot of $|\mu|=\log((r-1)C/2r)/\beta$ as a function of $r$ for different times from $10^3$ to $10^6$ in a model with $m=2$, $T=0.125$, $g(\varepsilon)=3\varepsilon^2$, averaged over 100 runs.  
The vertical line corresponds to the critical value of $r$ for the phase transition. The border between the Fermi-like and the Bose-like regimes is at $r=1$. }\label{p2_z}
\end{figure}

\begin{figure}
\begin{center}
\includegraphics[width=\columnwidth]{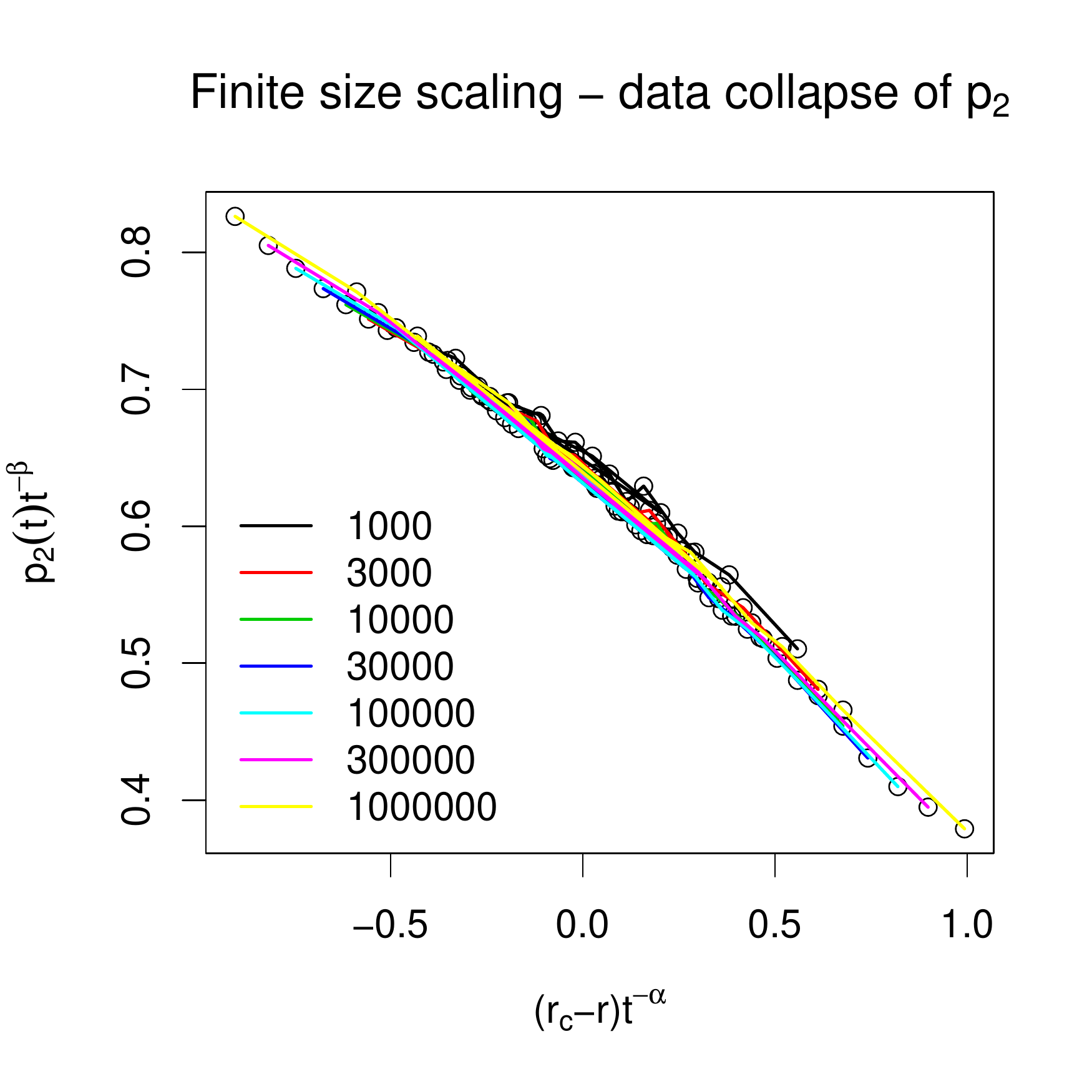}
\end{center} 
\caption{(Color online) Plot of data collapse for the finite size scaling of $p_2(t)$ across the transition  at $r_c\simeq 1.014$ in a model with $m=2$, $T=0.125$, $g(\varepsilon)=3\varepsilon^2$, averaged over 100 runs. The scaling parameters are $\alpha\simeq -1/12$ and $\beta\simeq -1/36$. 
}\label{fig_dc}
\end{figure}

From  the simulation of the model it is found that the maximum fraction of links on a single node decreases as an inverse power of time (Figure \ref{kmax}), implying  that there is no condensation of the links, i.e. there is no node with a finite fraction of all the links. Therefore $k(t)$ scales sublinearly and, by the argument above, $P_2$  should grow faster than $t/\log(t)$, i.e. almost linearly.  In order to check the validity of this conclusion we simulated the network model at low temperature.
By changing the rewiring rate $r$ 
we show that above the predicted phase transition $p_2$ becomes finite, i.e. $p_2$ doesn't show anymore finite size effects. The results are shown in Figure \ref{p2_max} and Figure \ref{p2_z}. Data collapse for the finite size scaling of $p_2$ is shown in Figure \ref{fig_dc}.

Therefore, in this system there is condensation of paths of length 2. This can be interpreted as an ``extended condensate'' of links on an infinitesimal fraction of nodes connected to a central node. This is still a Bose-Einstein condensation as described by the integral (\ref{boseint}), but different from the one observed in \cite{Bose,Weighted}, and strongly affected by the constraint of avoiding tadpoles and multilinks.

An upper bound for $p_2$ can be obtained from the self-consistent equation for the chemical potential $\mu$, or from the equivalent equation for conservation ot the total number of links.
In particular we  include in the conservation equation the contribution of the condensate of $2mt\bar{p}_2$ links and associated  with a value of $\varepsilon$ given by  $\varepsilon=0$. The bound on the size of the condensate is therefore
\beq
p_2<\bar{p}_2=\frac{r-r_c(\beta)}{r-1}\label{p2bound}
\eeq



\begin{figure}
\begin{center}
\includegraphics[width=\columnwidth]{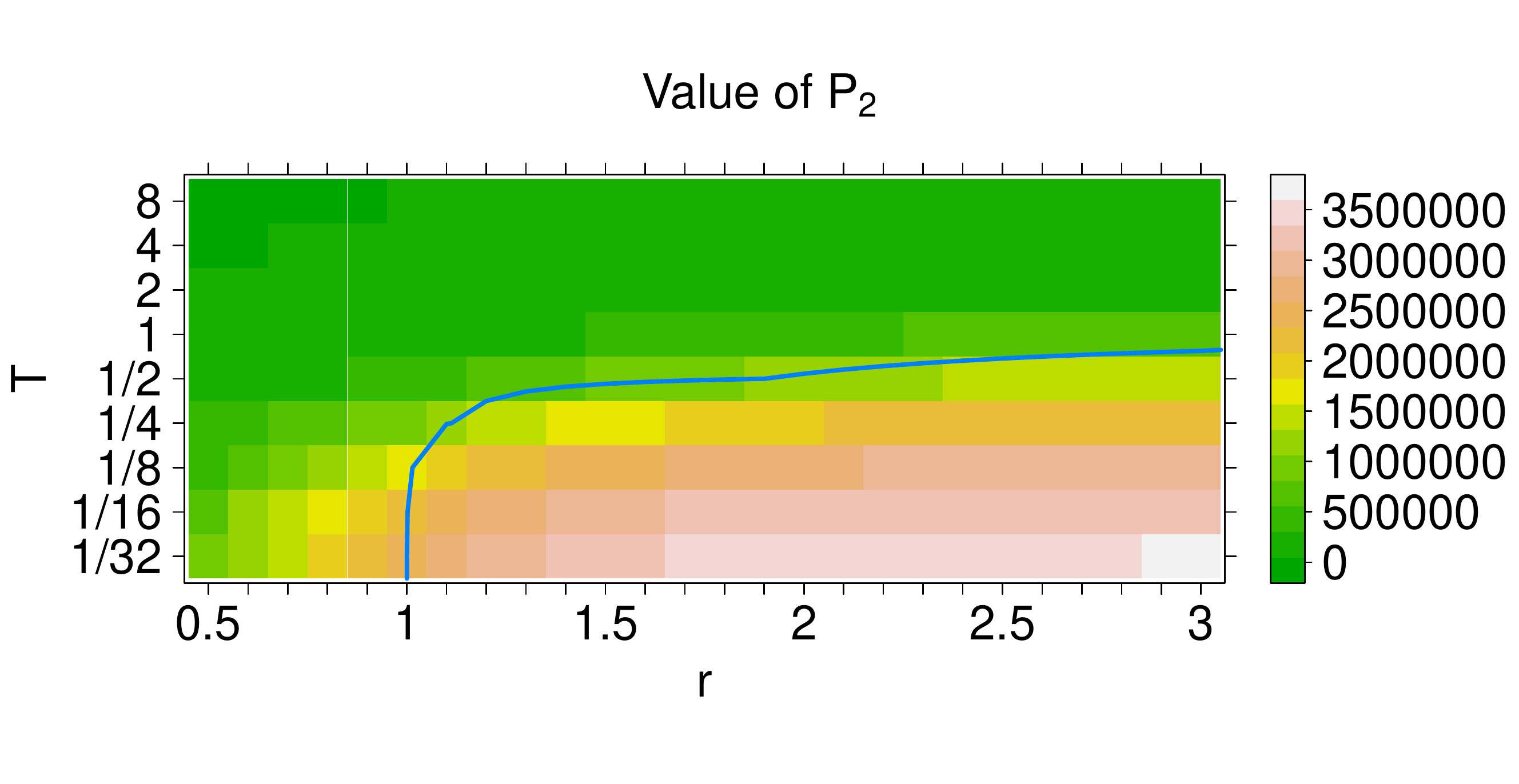}
\end{center} 
\caption{(Color online) Plot of $P_2$ at $t=10^6$ in a model with $m=2$, $g(\varepsilon)=3\varepsilon^2$, averaged over 100 runs.}\label{2d_p2}
\end{figure}

\begin{figure}
\begin{center}
\includegraphics[width=\columnwidth]{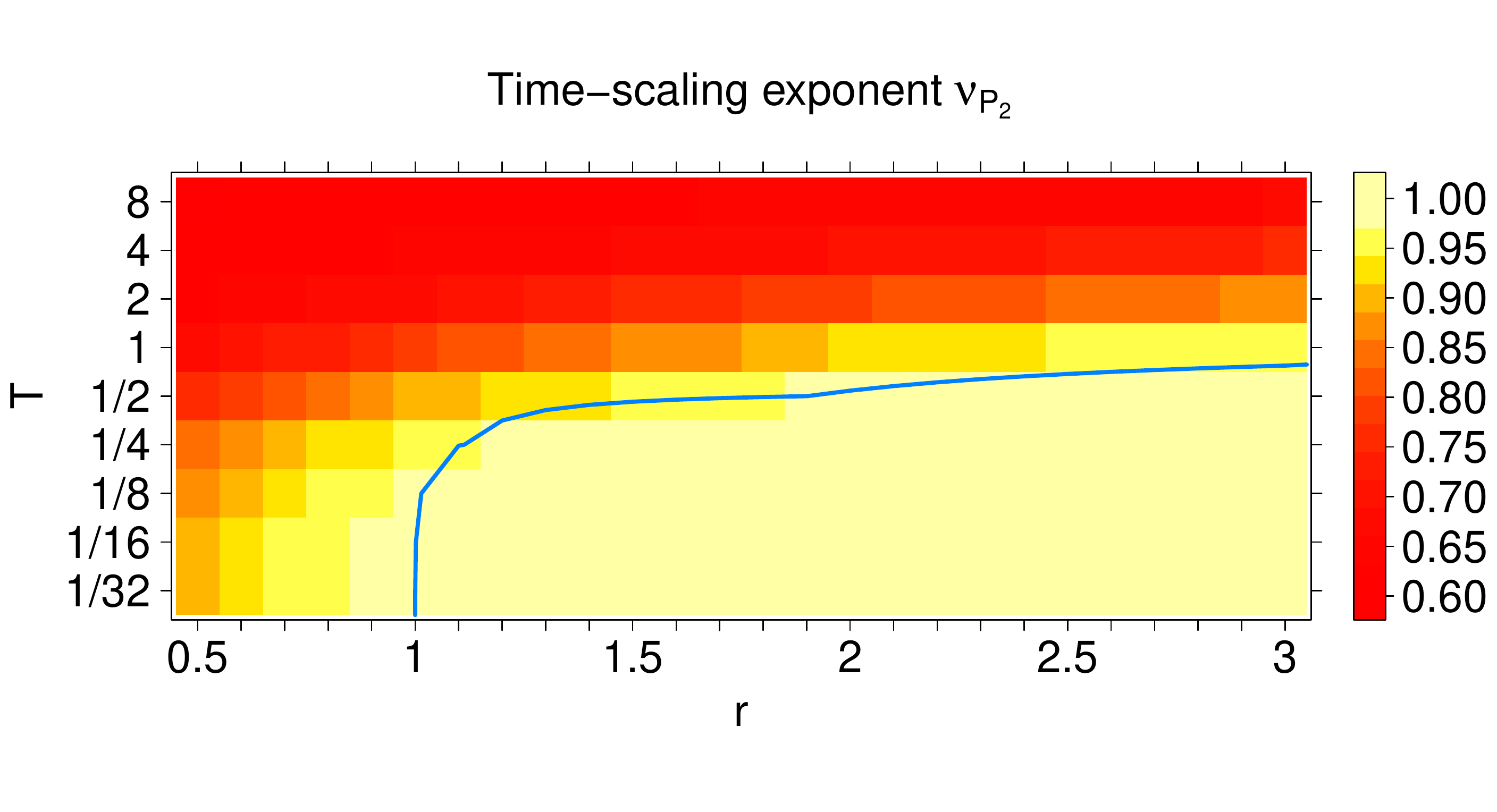}
\end{center} 
\caption{(Color online) Scaling exponent $\nu_{P_2}$ in the $r-T$ plane, computed between $t=10^4$ and $t=10^6$ in a model with $m=2$, $g(\varepsilon)=3\varepsilon^2$, averaged over 100 runs. The transition corresponds to $\nu_{P_2}=1$.  line represent the transition line predicted by Eq. (\ref{eqbrt}). }\label{scal2d_p2}
\end{figure}

The $r-T$ plot and the critical region for the condensate can be obtained from numerical simulations. 
Numerically we see that the scaling of $P_2$ in time between $t=10^3-10^7$ is well described by a power-law 
\beq
P_2(t)\sim t^{\nu_{P_2}}
\eeq
In Figure \ref{2d_p2} and 
Figure \ref{scal2d_p2} we plot respectively $P_2$ and its scaling exponent $\nu_{P_2}$. When the exponent $\nu_{P_2}$ reaches 1, a condensate is formed. 

We studied numerically also the variance of these quantities. The coefficient of variation of the exponent $\nu_{P_2}$ is at most 0.15 outside the condensate region and becomes close to 0 in the condensed phase.

\section{Phase transitions at $T=0$}
In order to obtain a better understanding of the condensation, we look at the special case of zero temperature.  

The phase transition can be described analitically for $T\rightarrow 0$. In this limit, the form of the distribution of energy is irrelevant, since links are removed deterministically from the nodes with higher energy and the only stochasticity is in the choice of the energies of the nodes.

\newcommand{\qe}{q_\varepsilon}

\newcommand{\qeb}{\bar{q}_\varepsilon}

For this reason, we can actually use the quantile $\qe$ of the energy distribution instead of the energy $\varepsilon$. The quantile is defined as $\qe=\int_0^\varepsilon d\varepsilon' g(\varepsilon')$. It plays the same role as $\varepsilon$, i.e. links with high $\qe$ tend to lose link more easily, but it also defines the probability that a random node has energy less than $\varepsilon$. 
We denote by $\qeb$ the quantile corresponding to the most energetic node that has still links. Nodes that are born with $\qe>\qeb(t)$ lose all their links in a very short time, while all nodes below $\qeb$ are equivalent except for their birth time. We also denote by $N^*(t)$ the number of nodes in the network that have some links at time $t$. Since these nodes correspond essentially to $\qe<\qeb(t)$, we have $N^*(t)\simeq t\qeb(t)$. Note that for simple graphs $N^*(t)\gtrsim t^{1/2}$, therefore $\qeb(t)\gtrsim t^{-1/2}$.

There are three phases in this system, depending on $r$.

\subsubsection*{Normal phase ($r<1$)}
In this phase, the rewiring process occurs at rate $r<1$. When this process occurs   the links attached to nodes with quantile  $\qe>\qeb$ are rewired while the nodes with quantile $\qe\leq \qeb$ are not affected by the rewiring process. Therefore the continuous equation  for the degree of a node $i$ with $\qe\leq \qeb$ is given by  
\beq
\frac{dk_i}{dt}=\frac{(1+r)k_i}{2t}.
\eeq 
For the above equation it follows that the degree  of these nodes increase as  $k_i(t)=m (t/t_i)^{(1+r)/2}$.
In this case the value of $\qeb$ is fixed by the conservation of the number of links,i.e.
\bea
\hspace*{-5mm}\sum_i k_i(t) \theta(\qeb-q_{\varepsilon_i})\simeq \qeb\int_1^t dt_i m (t/t_i)^{(1+r)/2} =2mt
\label{cons}
\eea
where $q_{\varepsilon_i}$ is the quantile of node $i$ and the Heaviside function $\theta (x)$ is defined as $\theta(x)=1$ if $x\geq 0$, otherwise $\theta(x)=0$. From Eq. $(\ref{cons})$ it follows that $\qeb=1-r$. Therefore the number of nodes in the network that have some links at time $t$ increases linearly with time, i.e. $N^*(t)\simeq(1-r)t$. The growth of $P_2$ is bounded from above by the sum of the links of the $k_{max}(t)$ most connected nodes, since the nodes connected to the central node of the condensate are at most $k_{max}(t)$: $P_2/2mt \leq
\int_1^{k_{max}(t)}dt_0\ k_{t_0}(t) =
\int_1^{mt^{(1+r)/2}}dt_0 (t/t_0)^{(1+r)/2}/2t\sim t^{-(1-r)^2/4}\rightarrow 0$,
so there is no condensate.

\subsubsection*{Condensed phase ($1<r<2$)}
In this phase, new nodes with $\qe>\qeb(t)$ lose all links almost immediately after being created and $r-1$ extra links are removed from the node at $\qeb(t)$. Existing nodes with $\qe=\qeb(t)$ have on average $2mt/N^*(t)$ links and they lose $m(r-1)(1-\qeb(t))+mr\qeb(t)=m(r-1+\qeb(t))$ links per unit time, so these nodes disappear from the connected component at a mean rate $(r-1+\qeb(t))N^*(t)/2t$. When they lose all their link, $\qeb(t)$ becomes the quantile of the next node, which is on average $\qeb(t)(N^*(t)-1)/N^*(t)$, so the jump in $\qeb(t)$ is $-\qeb(t)/N^*(t)$. Taking all this into account, we can write an average equation for $\qeb(t)$ by multiplying the jump in $\qeb(t)$ by the rate at which nodes disappear from the connected component, valid also for $r<1$:
\beq
\frac{d\qeb}{dt}=-\qeb\frac{r-1+\qeb}{2t}
\eeq
therefore
\beq
\qeb(t)=\frac{1-r}{1-\frac{r-1+\qeb(0)}{\qeb(0)}t^{-(1-r)/2}}
\label{fs_n}
\eeq
For large times and $r>1$, we have therefore 
\beq
\qeb(t)\sim t^{-\frac{r-1}{2}}\quad,\quad N^*(t)\sim t^{\frac{3-r}{2}}\label{eqqn}
\eeq
so nodes of positive degree represent just an infinitesimal fraction of the network. The constraints on multilinks then imply that $k(t)\lesssim N^*(t)<t$, so there is no condensation of the links on a single node. 

\begin{figure}
\begin{center}
\includegraphics[width=\columnwidth]{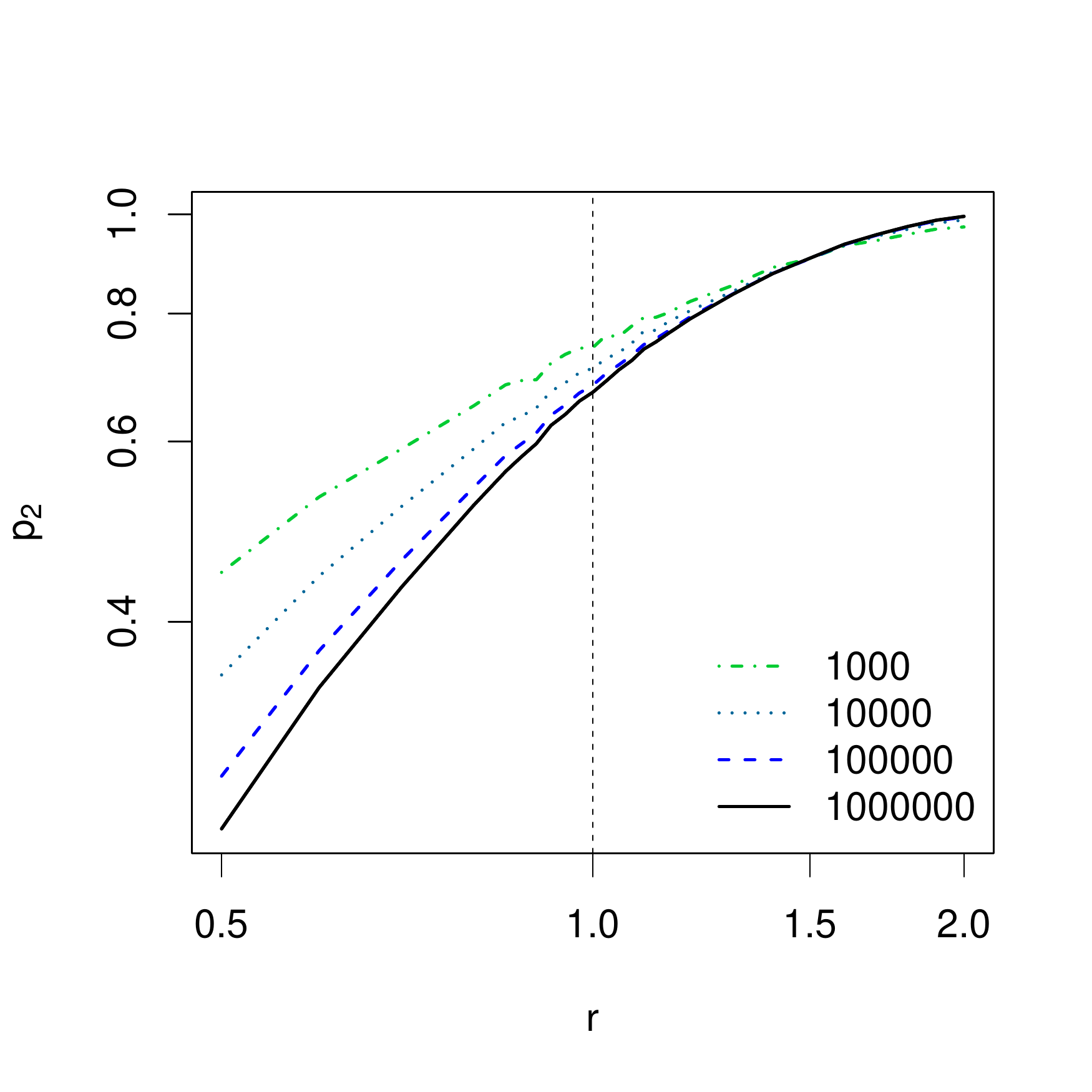}
\end{center} 
\caption{(Color online) Plot of $p_2$ as a function of $r$ in a model with $m=2$, $T=0$ for different times from $10^3$ to $10^6$, averaged over 100 runs.}\label{p2t0_max}
\end{figure}

At time $t_0$, the node with minimum energy should have $\qe\sim 1/t_0$. This node survives and increases $P_2$ until a time $t'$ such that $\qeb(t')=1/t_0$, i.e. $t'\sim t_0^{2/(r-1)}$. 
We can neglect the degrees of individual nodes since they represent an infinitesimal fraction of links. In this approximation, $P_2$ is the total connectivity of the nearest neighbours, 
so it grows due to the attachment of new links at rate $m\sum_{NN}\Pi_+=m\frac{P_2}{2mt}$ and the rewiring of existing links at rate $mr\sum_{NN}\Pi_+(1-p^{(self)}-p^{(mult)})$. The total probability of generating selflinks or multiple links is at most $p^{(self)}+p^{(mult)}\leq P_2/2mt$.   
Therefore, the evolution of $P_2$ satisfies
\beq
\frac{dP_2}{dt}\gtrsim \frac{P_2}{2t}+\frac{rP_2}{2t}\left(1-\frac{P_2}{2mt}\right)
\eeq 
Solving this equation, we obtain 
\beq
P_2(t)\gtrsim 2mt\left(\frac{P_2(t_0)\left(t/t_0\right)^{\frac{r-1}{2}}}{2mt_0+\frac{r}{r-1} P_2(t_0)\left(\left(t/t_0\right)^{\frac{r-1}{2}}-1\right)}\right)
\eeq 
and with an initial condition 
$P_2(t_0)$ which is larger than $2mt_0/N^*(t_0)$, it is easy to show 
that $p_2(t')\sim O(r-1)$ at least, independently of $t'$. Therefore, there is a condensation of $P_2$. This is confirmed by numerical simulations  (Figure \ref{p2t0_max}). 

We have shown explicitly that  above the critical rewiring rate $r_c=1$ there is a Bose-Einstein phase transition to a condensate phase. This is consistent with the prediction $r_c(\beta=\infty)=1$ from Eq. (\ref{eqbrt}). In particular, there is condensation of paths of length 2 on one of the nodes of lower energy. Note that as in the BB model \cite{Bose,ferretti}, the condensate is dynamical, i.e. new nodes of lower energy can overcome the old condensate and become new condensates themselves. 

\begin{figure}
\begin{center}
\includegraphics[width=\columnwidth]{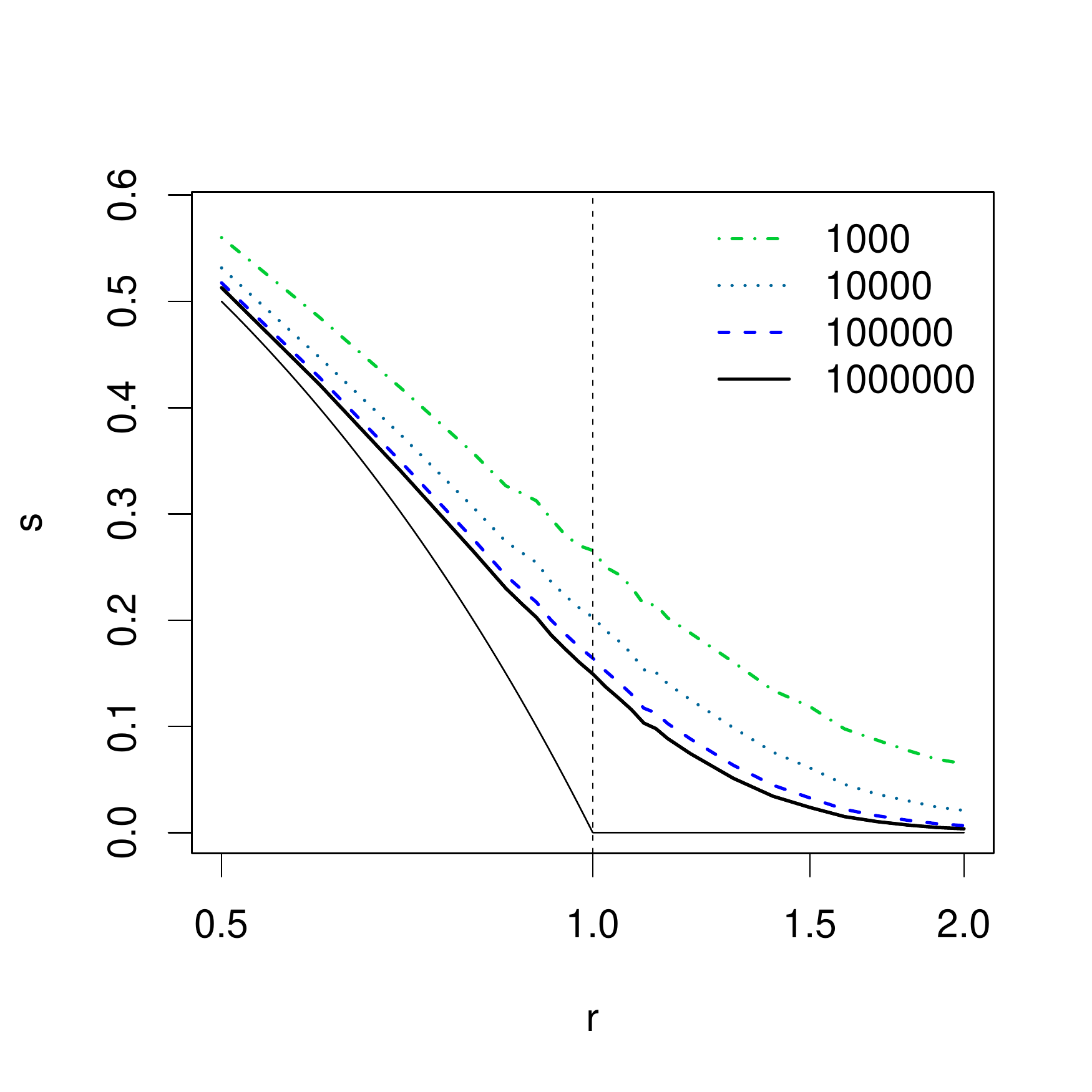}
\end{center} 
\caption{(Color online) Plot of $s$ as a function of $r$ in a model with $m=2$, $T=0$ for different times from $10^3$ to $10^6$, averaged over 100 runs. The bottom line is the predicted asymptotic value of $s$.}\label{s_t0}
\end{figure}

\begin{figure}
\begin{center}
\includegraphics[width=\columnwidth]{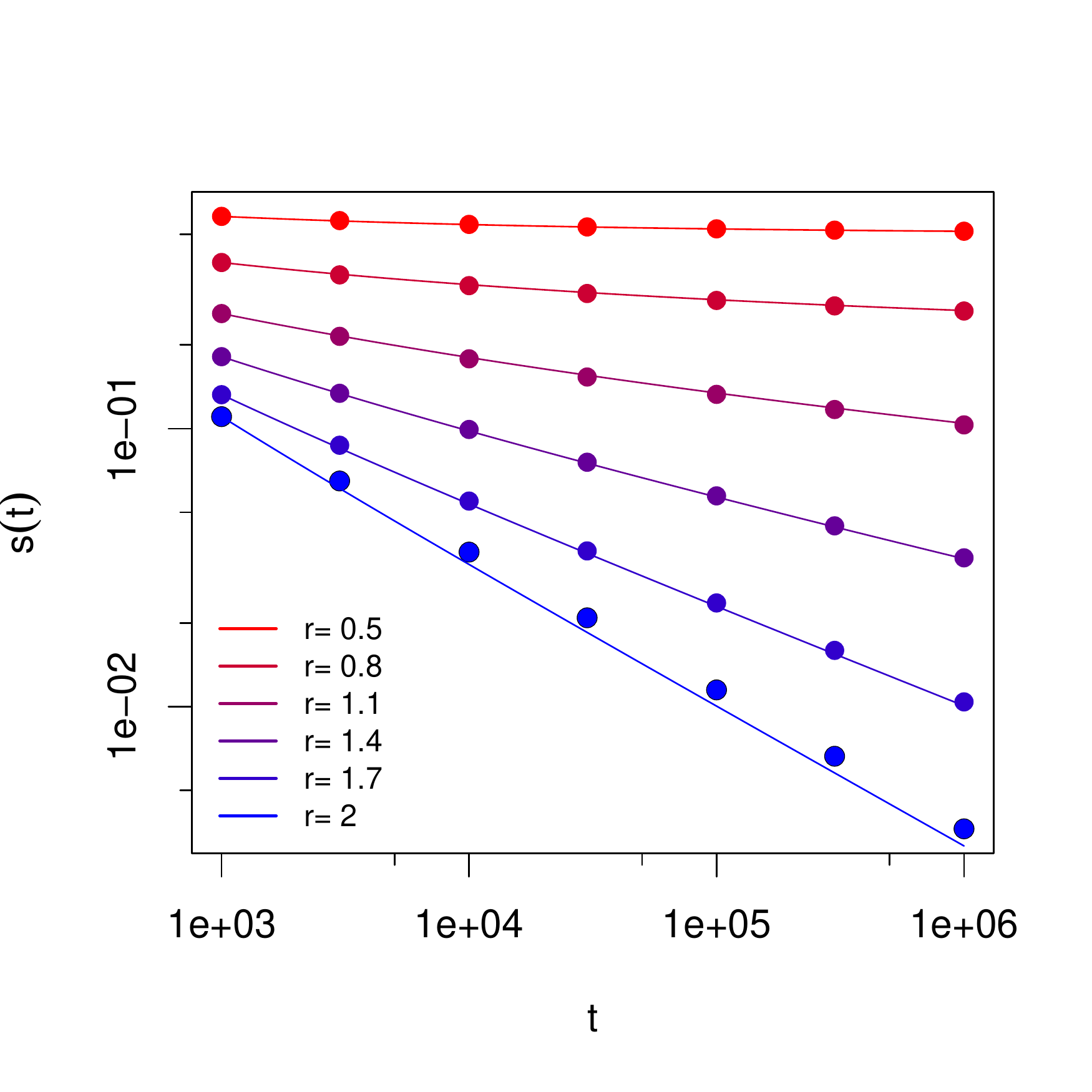}
\end{center} 
\caption{(Color online) Plot of $s(t)$ for different values of $r$ in a model with $m=2$, $T=0$. The points represent the average over 100 runs of simulations while the continuous lines are the analytical predictions. }\label{s_t0_fit}
\end{figure}

Note that at $r=1$ there is also a second-order phase transition in the size of the connected component $s$, since $s(t)=\qeb(t)=N^*(t)/t$. The predicted asymptotic value is 
\beq
s(+\infty)=\begin{cases}(1-r) & \mathrm{for}\ r<1 \\ 0   & \mathrm{for}\ r\geq 1 \end{cases}
\eeq 
Numerical results for this transition are presented in Figures \ref{s_t0}, \ref{s_t0_fit}. Simulations confirm the finite size scaling (\ref{fs_n}). 

\subsubsection*{Complete graph ($r>2$)}
In this phase we see from Eq. (\ref{eqqn}) that $\qeb(t)\sim t^{-1/2}$ and $N^*(t)\sim t^{1/2}$, i.e. the network resembles a complete graph (all nodes are connected to almost all the other nodes) plus a large fraction of isolated nodes of degree 0. In this case, the neighbours of all nodes of positive degree form almost the entire graph, therefore $p_2\simeq 1$.


\section{Topological transitions}\label{top}

From the above analysis at $T=0$, the phase transition does not only appear as condensation of links/paths, but it changes also other aspects of the topology of the network. In particular, the scaling of the size of the connected component changes from linear to sublinear at $r=1$. Furthermore, there is another transition at $r=2$ that is topological in nature, since the network becomes a complete graph for higher $r$. 

The presence of such a transition at finite temperature can be guessed by considering the behaviour at large $r$. We take the limit $r\rightarrow\infty$ at constant $T$ for a network of fixed size $t$. In this limit, the network collapses to the stable state under rewiring, that is a complete graph of size $(mt)^{1/2}$ plus isolated nodes. However this argument does not imply the existence of true phase transition in the thermodynamic limit, since the limit of large $r$ for the asymptotic dynamic (that is, taking the limit $r\rightarrow\infty$ after taking the limit $t\rightarrow\infty$ first)  is more complicated. 

In this section we study numerically several topological quantities and their transitions. We show that there is a transition in the scaling of the connected component across the BE transition. Moreover, there is a second transition for the degeneracy of the network, located inside the condensed phase.

\subsection{Scaling of the connected component}

The first topological observable is the scaling of the connected component $S(t)$. In this system, due to rewiring, there is only one large connected component, so $S(t)=N^*(t)$, i.e. the fraction of connected nodes (that is, nodes with at least one link). Numerically, its scaling in time is given by
\beq
S(t)\sim t^{\nu_S}
\eeq
The quantity that we measure is $s(t)=S(t)/t$, the size of the connected component divided by the size of the network. Since $s(t)\sim t^{\nu_S-1}$, the fraction $s$ is asymptotically finite only if $\nu_S=1$.  

\begin{figure}
\begin{center}
\includegraphics[width=\columnwidth]{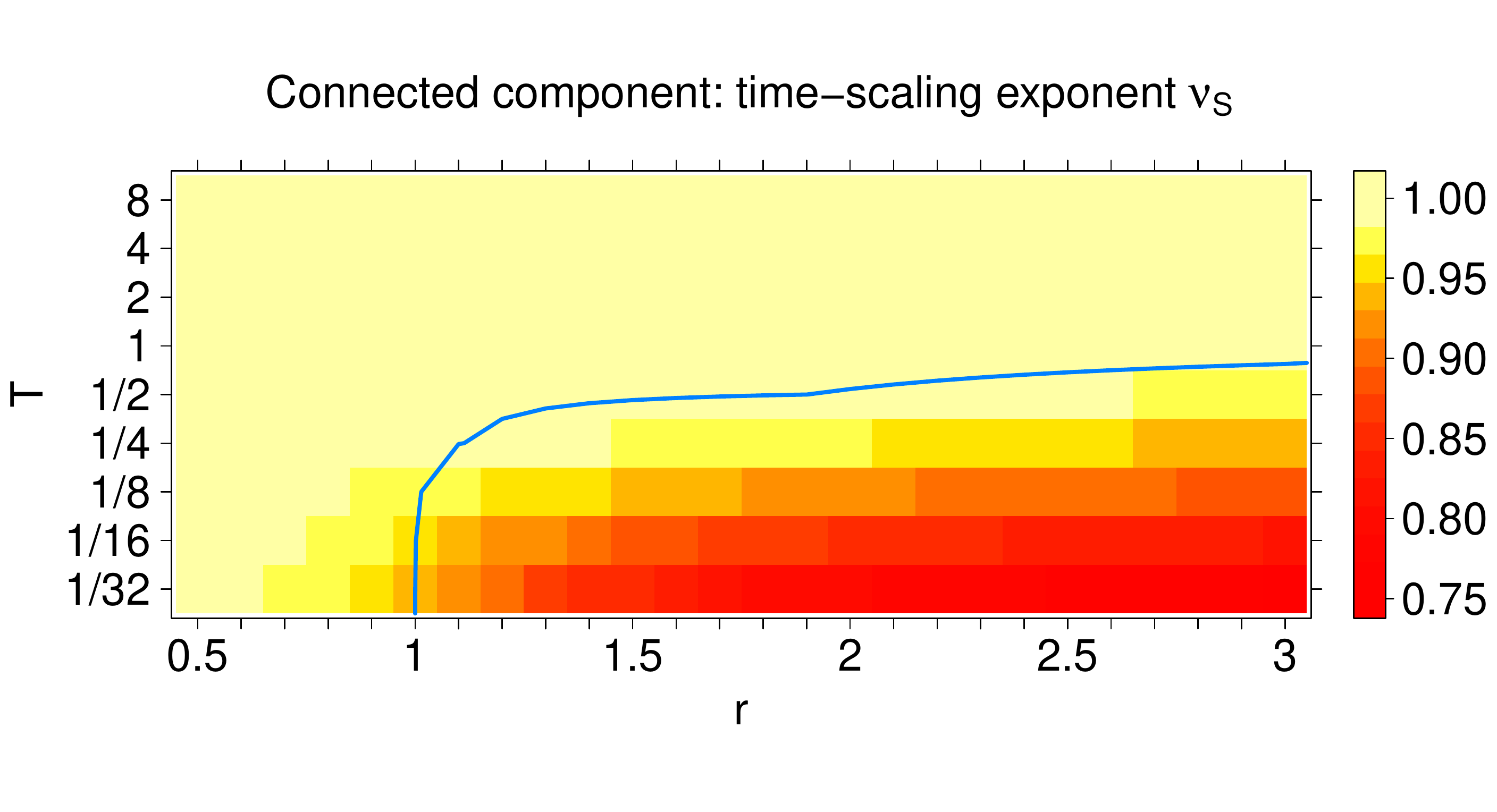}
\end{center} 
\caption{(Color online) Scaling exponent of the number of connected nodes $\nu_S$ in the $r-T$ plane, computed between $t=10^4$ and $t=10^6$ in a model with $m=2$, $g(\varepsilon)=3\varepsilon^2$, averaged over 100 runs. The transition corresponds to $\nu_S<1$.}\label{scal2d_nonzero}
\end{figure}

In the thermodynamic limit, $s$ is always positive outside the condensed phase. However, numerical simulations in Figure \ref{scal2d_nonzero} 
show that $\nu_S<1$ in the condensed phase, i.e. $s(t)\rightarrow 0$ in the thermodynamic limit. Since $s(t\rightarrow\infty)$ has a finite limit near the boundary of the condensed phase, there is a transition of $s$ across the condensate transition. This is in agreement with the transition at $T=0$, where we showed analytically that $s(t)\sim 1-r$ for $r<1$ and $s(t)\sim t^{-(r-1)/2}$ for $r>1$.  

\begin{figure}
\begin{center}
\includegraphics[width=\columnwidth]{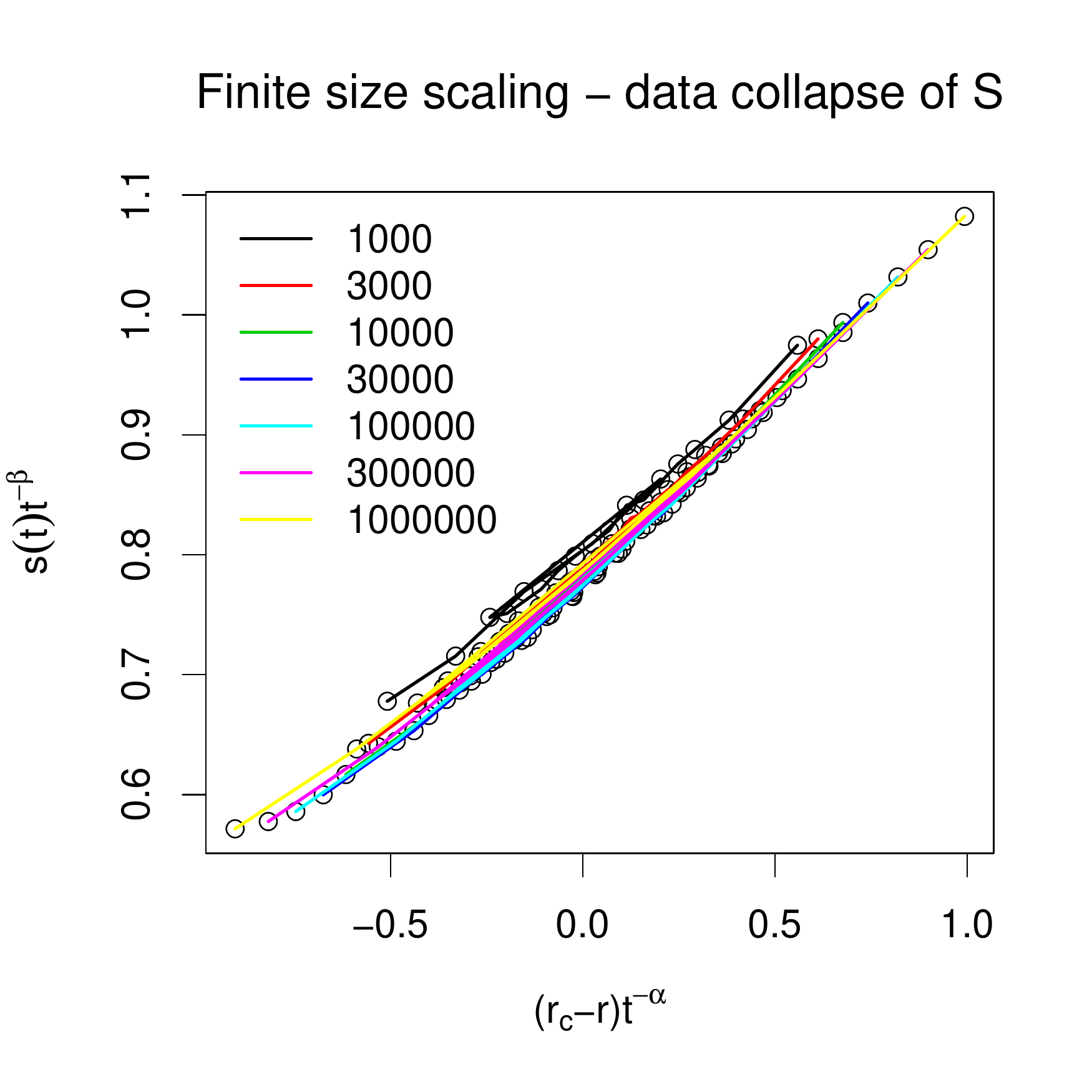}
\end{center} 
\caption{(Color online) Plot of data collapse for the finite size scaling of $s(t)$ across the transition  at $r_c\simeq 1.014$ in a model with $m=2$, $T=0.125$, $g(\varepsilon)=3\varepsilon^2$, averaged over 100 runs. The scaling parameters are $\alpha\simeq -1/12$ and $\beta\simeq -1/36$. 
}\label{fig_dc_s}
\end{figure}

The coefficient of variation for the scaling exponent $\nu_S$ is small ($<0.25$ outside the condensed phase and $<0.1$ in the condensed phase, from 100 runs). Data collapse for the finite size scaling of $S$ is shown in Figure \ref{fig_dc_s}.

\subsection{Clustering of the condensate}

In the Barabasi-Albert and the Bianconi-Barabasi fitness model, there is no clustering in the thermodynamic limit. In fact, the scaling of the clustering coefficient in time is given by $C(t)\sim t^{\nu_C}$ with $\nu_C<0$ \cite{ferretti2012features}. In contrast, in our model there is a transition to finite clustering at $T=0$, $r\geq 2$. We would like to explore the behaviour of the clustering coefficient at finite temperature.

In the condensed phase, the central node of the condensate and all its neighbours form a strongly connected subnetwork. In fact, numerical simulations show that in this phase, the fraction of links in this subnetwork scales linearly with the network size. This suggests that there could be a transition also in the clustering coefficient of the central node.

However, a simple argument shows that it is difficult to have finite clustering around the condensate node: the clustering coefficient of the condensate node cannot be asymptotically finite unless the node degree grows as $t^{1/2}$ or slower. 
In fact, if we assume that the degree of the condensate node grows like $t^{\beta'}$, since the number of possible triangles grows quadratically with the degree while the links grows linearly in time, an upper bound on its clustering is $C(t)\lesssim t/t^{2\beta'}$ that is finite only if $\beta'\leq 1/2$.

For low rewiring rates we have $\beta'>1/2$, so it is possible to have a finite clustering only if the rewiring rate is high enough. In particular, by joining Eqs. (\ref{keq}) and (\ref{p2bound}), we have $k(t)\lesssim t^{1/2}$ only if $r- r_c\geq\langle\xi_{NP_2}\rangle^{-1}$. 
Since $\langle\xi_{NP_2}\rangle\leq \langle\xi\rangle$, we obtain that a necessary condition for finite clustering at $t\rightarrow\infty$ is $r\geq r_c+\langle\xi\rangle^{-1}$. 
This means that any further transition to finite clustering, if possible, should occur well inside the condensed phase. 
The above bounds are consistent with the results at zero temperature. For $T=0$, $\langle\xi\rangle_{T=0}=+\infty$, so the transition should occur at $r\geq r_c=1$. Actually, we know that $\langle\xi_{NP_2}\rangle_{T=0}=1$ so the stricter bound should be $r\geq r_c+1=2$. In fact, the transition to finite clustering at $T=0$ corresponds to the transition to an almost complete graph and occurs precisely at $r=2$.

\begin{figure}
\begin{center}
\includegraphics[width=\columnwidth]{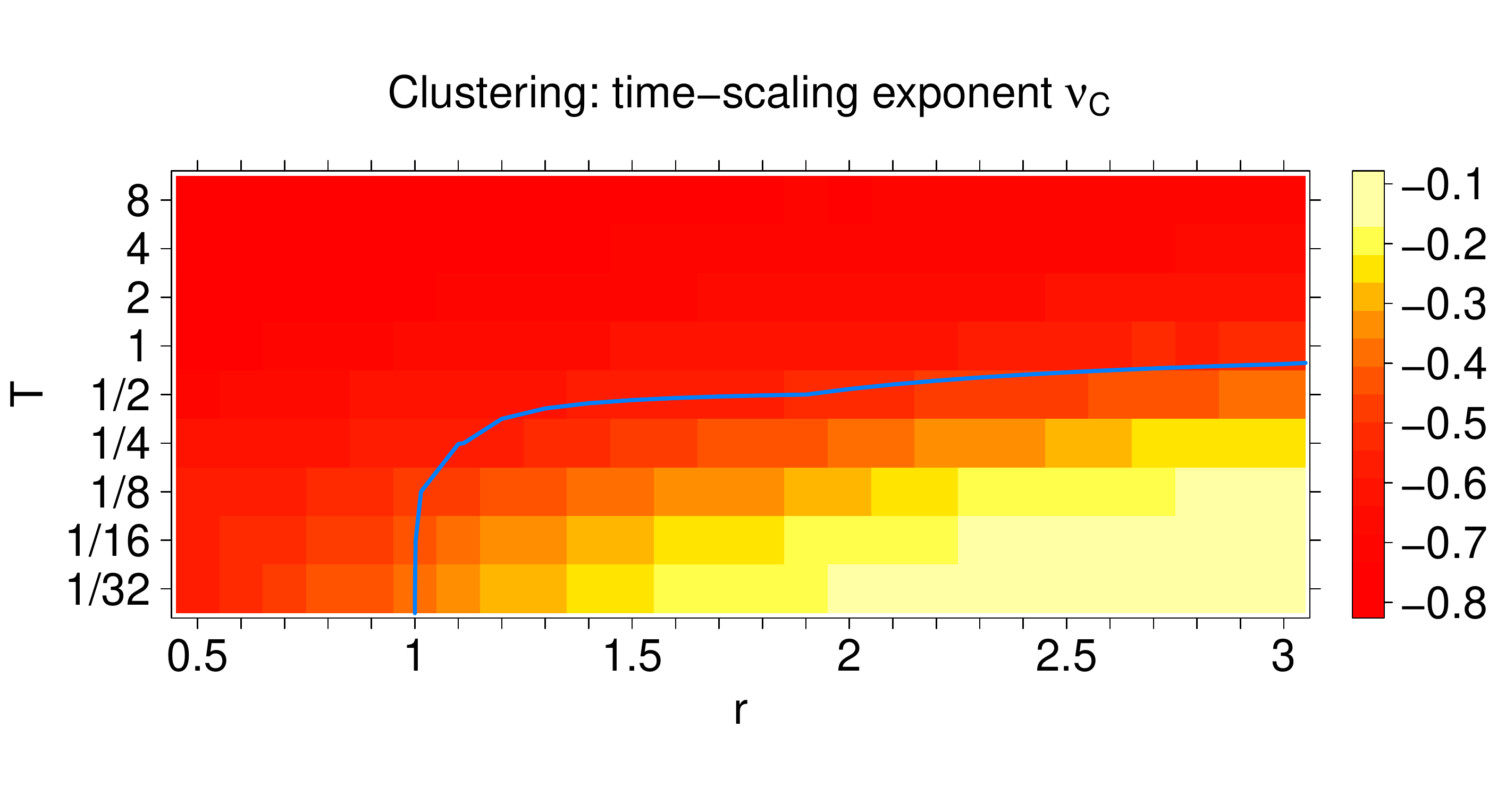}
\end{center} 
\caption{(Color online) Scaling exponent of the clustering coefficient $\nu_C$ in the $r-T$ plane, computed between $t=10^4$ and $t=10^6$ in a model with $m=2$, $g(\varepsilon)=3\varepsilon^2$, averaged over 100 runs. The transition would correspond to $\nu_{C}=0$.}\label{scal2d_clust}
\end{figure}

Numerical simulations for the clustering coefficient of the central node of the condensate at $T>0$ are shown in Figure 
\ref{scal2d_clust}. There appears to be a region of finite clustering for $r\geq 4$ (not shown), however this could be a finite-size effect related to the limit of large $r$ at fixed size discussed above. The coefficient of variation for the  exponent $\nu_C$ is low ($<0.15$) outside the condensate region, but it grows to $0.3$ and more in the condensed phase where the scaling is closer to 0. The presence of a transition for the clustering coefficient at $T>0$ is therefore unclear.

\subsection{$K$-cores and degeneracy of the network}

Another quantity that could describe the strong linkage of the condensate is the degeneracy, i.e. the largest value of $K$ such that the network has a non-empty $K$-core. $K$-cores are defined as the maximal connected subset of nodes such that each node has at least $K$ internal links in the induced subnetwork. The minimum number of nodes and links inside a $K$-core of a simple graph are $K+1$ and $K(K+1)/2$ respectively, so the maximum $K$-core for a network with $t$ nodes and $l$ links scales as $K_{max}\sim \min(l^{1/2},t)$. In our model $l=mt$ so $K_{max}\sim t^{1/2}$.  

In the thermodynamic limit, we consider the ratio of the degeneracy to its maximum $K/K_{max}$. This is a constant if $K\sim t^{1/2}$, while it goes to zero if $K$ grows slower than $t^{1/2}$. We would like to see if there is a transition associated to this quantity, or equivalently to the scaling of $K$. Assuming a power-law scaling
\beq
K(t)\sim t^{\nu_K}
\eeq
in agreement with simulations, we characterize a transition to maximal degeneracy when $\nu_K=1/2$.

\begin{figure}
\begin{center}
\includegraphics[width=\columnwidth]{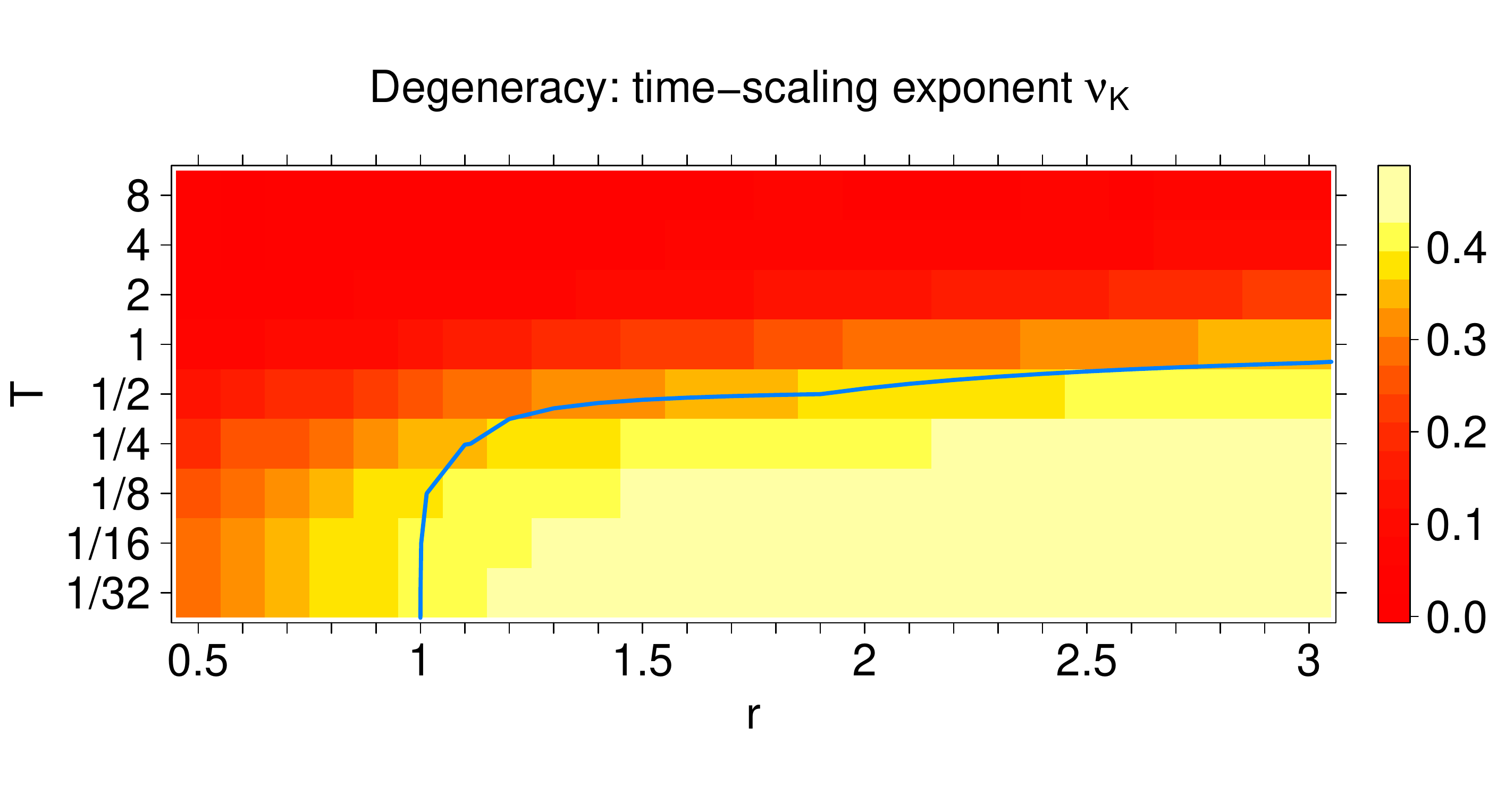}
\end{center} 
\caption{(Color online) Scaling exponent of the degeneracy $\nu_K$ in the $r-T$ plane, computed between $t=10^4$ and $t=10^6$ in a model with $m=2$, $g(\varepsilon)=3\varepsilon^2$, averaged over 100 runs. The transition corresponds to $\nu_K=1/2$.}\label{scal2d_deg}
\end{figure}

Numerical results for the degeneracy in our model are shown in Figure 
\ref{scal2d_deg}. The coefficient of variation for the scaling of degeneracy can reach 1.5 at high temperatures and low rewiring, but it drops to $<0.2$ in the region near the condensation transition and to $<0.1$ in the condensed phase. 

Interestingly, there appears to be a transition to a low-temperature/high-rewiring phase where the degeneracy grows almost with the maximum scaling $t^{1/2}$. This transition is topological in nature, since it depends on the global topology of the network and not only on the distribution of links. The simulations show that this phase transition is located well inside the condensed phase, therefore the model has at least three different phases in the $r-T$ plane. \newline

%
%
%

\section{Discussion and conclusions}

In this paper we have studied a dynamical network model in which the rewring of the links occurs according to a preferential attachment rule modulated by the effect of a negative fitness $\xi_i$ associated to each node which describe the ability of the node $i$ to loose links.
In this way the network evolves toward a highly optimized structure.
The signature of the strong optimization that can be reached in this networks is given by the possible values of the condensate fraction of links $\bar{p}_2(\beta)>0$, which can reach 100\% for some parameters. 

We provide a mean-field solution of the model above the condensation phase transition and we have shown by numerical simulations and analytical arguments that below the phase transition the model is in a structurally different phase with a finite fraction of the links on the nearest neighbours of the condensed node.  
The major characteristics of the present model are outlined below.

First, the present model includes the role of a negative fitness of nodes $\xi$, determining the rate at which links are rewired from the nodes,  and provides a new condensation phase transition with respect to the Bianconi-Barabasi fitness model that only include a positive fitness of the links. In this sense, this model is based on the removal of the most energetic links/particles and resembles more the mechanism of evaporative cooling for experimental BE condensates.

Secondly,  in the present model the quantity that undergoes the condensation transition is the number of paths of length $2$ from the  node with smaller negative fitness, or equivalently the number of links on its nearest neighbours.  This is to our knowledge the first example of the condensation of such structure in evolving network models. This phenomena is a consequence of the fact that in our model, when we perform a rewiring of a link, we do not allow for selflinks or multilinks, thereby enforcing the topological constraint of a simple graph.

Furthermore, this model has a richer phase space with two parameters, the temperature $T$ and the rewiring rate $r$, and multiple transitions. The BE phase transition changes also the scaling of the giant component, while at $T=0$ there is another transition to a complete graph at $r=2$ and for $T>0$ we have a similar transition to the maximal scaling of the degeneracy of the network. 
These phase transitions are located well inside the condensed phase, therefore the model shows a quite rich structure with at least three different phases in the $r-T$ plane.



\acknowledgments

LF acknowledges support from the grant ANR-12-JSV7-0007 from ANR (France).

\end{document}